\documentclass[preprint,12pt]{elsarticle}
\usepackage{lineno,hyperref}
\usepackage{amssymb}
\usepackage{subfigure}
\usepackage{caption}
\usepackage{epsfig}
\usepackage{graphicx}
\usepackage{dcolumn}
\usepackage{bm}
\usepackage{times}
\usepackage{xcolor}
\usepackage[percent]{overpic}
\usepackage{hyperref}
\usepackage[hyphenbreaks]{breakurl}
\usepackage{multirow}
\usepackage{slashbox}
\usepackage{url}
\usepackage{float}
\usepackage{textcomp}
\usepackage{amsmath}

\begin{document}

\begin{frontmatter}

\title{Effects of heterogeneous self-protection awareness on resource-epidemic coevolution dynamics}


\author[address1]{Xiaolong Chen}

\author[address1]{Kai Gong}

\author[address2]{Ruijie Wang\corref{mycorrespondingauthor}}
\cortext[mycorrespondingauthor]{Corresponding authors}
\ead{ruijiewang001@163.com}

\author[address3,address4,address5]{Shimin Cai}
\author[address6]{Wei Wang}
\ead{wwzqbx@hotmail.com}
\address[address1]{School of Economic Information Engineering, Southwestern University of Finance and Economics, Chengdu 611130, China}
\address[address2]{A Ba Teachers University, A Ba 623002, China}
\address[address3]{School of Computer Science and Engineering, University of Electronic Science and Technology of China, Chengdu 610054, China}
\address[address4]{Institute of Fundamental and Frontier Sciences, University of Electronic Science and Technology of China, Chengdu 610054, China}
\address[address5]{Big Data Research Center, University of Electronic Science and Technology of China, Chengdu 610054,China}
\address[address6]{Cybersecurity Research Institute, Sichuan University, Chengdu 610065, China}

\begin{abstract}
  Recent studies have demonstrated that the allocation of individual resources has a significant influence on the dynamics of epidemic spreading. In the real scenario, individuals have a different level of awareness for self-protection when facing the outbreak of an epidemic. To investigate the effects of the heterogeneous self-awareness distribution on the epidemic dynamics, we propose a resource-epidemic coevolution model in this paper. We first study the effects of the heterogeneous distributions of node degree and self-awareness on the epidemic dynamics on artificial networks. Through extensive simulations, we find that the heterogeneity of self-awareness distribution suppresses the outbreak of an epidemic, and the heterogeneity of degree distribution enhances the epidemic spreading. Next, we study how the correlation between node degree and self-awareness affects the epidemic dynamics. The results reveal that when the correlation is positive, the heterogeneity of self-awareness restrains the epidemic spreading. While, when there is a significant negative correlation, strong heterogeneous or strong homogeneous distribution of the self-awareness is not conducive for disease suppression. We find an optimal heterogeneity of self-awareness, at which the disease can be suppressed to the most extent. Further research shows that the epidemic threshold increases monotonously when the correlation changes from most negative to most positive, and a critical value of the correlation coefficient is found. When the coefficient is below the critical value, an optimal heterogeneity of self-awareness exists; otherwise, the epidemic threshold decreases monotonously with the decline of the self-awareness heterogeneity. At last, we verify the results on four typical real-world networks and find that the results on the real-world networks are consistent with those on the artificial network.
\end{abstract}


\end{frontmatter}

\section{Introduction} \label{sec:intro}
Resources, such as funds, medical and protective equipment, play a vital role in
constraining or mitigating the outbreak of an epidemic.
However, the pandemics always announced themselves with a sudden explosion of cases, inducing a severe shortage of public resources
\cite{zingg2005impact,tsai2008resource}. Examples
including the SARS coronavirus in 2003~\cite{syed2010developing} and Ebola virus
in 2014~\cite{gostin2015retrospective} etc..
At the end of 2019, a new type of coronavirus called COVID-19
broke out in Wuhan, China, and it quickly spreads around the world.
By the end of April 2020, more than three million cases worldwide have been officially reported~\cite{worldcoronavirus}. The increasing demands for protection and treatment have led to a severe shortage of public resources \cite{COVID-19shortage}.

Facing the shortage of public resources, the topic of
optimal resource allocation in suppressing disease spreading has aroused extensive attention from varies communities \cite{preciado2013optimal,lokhov2017optimal,li2020functional,zhang2020optimal,zhang2019suppressing,guo2020spatial}. For example, Andrey et al.~\cite{lokhov2017optimal}
solved the problem of optimal deployment of limited resources
by studying the interplay between network topology
and spreading dynamics. Based on a scalable dynamic message-passing approach \cite{li2020functional}, they got the optimal distribution of available resources and demonstrated the universality of the method on a variety of real-world examples. Preciado et al.~\cite{preciado2013optimal} researched the problem of how to optimally allocate the vaccination resources
on complex networks. Through a convex framework, they found the cost-optimal distribution of
the resources.
Nicholas et al. \cite{watkins2016optimal} developed a framework
to find an optimal strategy of resource allocation to eliminate
one of the epidemics when two competitive epidemics spreading on a bilayer network.
Besides, Nowzari et al. ~\cite{nowzari2015optimal} studied the problem of containing an initial epidemic outbreak under
budget constraints based on the analysis of a generalized epidemic model over arbitrary directed
graphs with heterogeneous nodes.
Chen et al.~\cite{chen2017optimal} studied the problem of optimally allocate the limited
resources to minimize the prevalence. They solved the problem
under the premise of the positive correlation between node degree and resource own by each node.

The previous works only considered the optimization of public resources and
addressed the problem from a mathematical perspective. However, during the
outbreak of epidemics, public resources such as medical staff, protective equipment are in a severe shortage. Usually, public resources are allocated preferentially to meet the needs of
severe and critical patients \cite{wu2020characteristics}. The treatment and
protection of mild or susceptible individuals mainly depend on the accumulation of individual
resources and the support of resources among individuals.
The topic that the influence of individual
resource on the epidemic spreading
has attracted wide attention in physic community \cite{bottcher2015disease,chen2018suppressing,chenoptimal}. For
instance, B{\"o}ttcher \cite{bottcher2015disease} considered that the healthy individuals could
contribute resources during an outbreak of an epidemic. By studying the coevolution of the resource
and disease, they found an ``explosive" increase of infected nodes induced by resource constraints.
Chen et al.~\cite{chen2018suppressing} studied the interplay between resource allocation
and disease spreading on top of multiplex networks, and found a hybrid phase
transition. In this multiplex network framework,
they further investigated the impact of preferential resource allocation on the
dynamics of epidemic spreading~\cite{chenoptimal}.

In real scenario, the diffusion of information/awareness can change human behaviors, such
as wearing masks or staying at home to reduce the frequency of face-to-face contact~\cite{zhan2018coupling,kabir2019analysis1}.
The interplay between information/awarenss diffusion and epidemic
spreading is another topic that has inspired a wide range of research by scholars
\cite{funk2010endemic,wang2015coupled,kabir2019effect,kabir2020impact}. For example,
Granell et al. \cite{granell2013dynamical} studied the dynamical interplay
between the epidemic spreading and information diffusion on top of multiplex networks.
Funk et al. \cite{funk2009spread}
studied the coevolution of information and disease on well-mixed populations and lattices, and
found that in a well-mixed population, the information about the disease
can suppress the disease. Kabir et al.~\cite{kabir2019analysis2} proposed a two-layer susceptible-infected-recovered/unaware-aware (SIR-UA) epidemic model to investigate
the impact of awareness on epidemic spreading on top of different heterogeneous networks.
Moreover, Kabir et al. \cite{kabir2019impact} further studied the effects of
awareness on the epidemic spreading based on a metapopulation model combined with a SIS-UA
(susceptible-infected-susceptible-unaware-aware) epidemic model.

Information or awareness can also change the individuals'
willingness of resource donation ~\cite{kan2017effects}.
In reality, the susceptible individuals would weigh whether to help others or protect themselves
when they are aware of the disease.
Since individuals are in various circumstances and have different personalities,
there is a heterogeneous distribution of awareness for self-protection.
For example, a cautious person is more likely to reserve resources
for self-protection than a generous person during an outbreak,
or a person who has received help from others will have a stronger
willingness to donate resources than others.
There remains a question that how the heterogeneous distribution of
awareness for self-protection influence the
epidemic dynamics. To answer this question, a resource-epidemic coevolution model is proposed,
which is based on the following assumption:
Namely, each susceptible individual has both a probability of resource donation
and a certain level of self-awareness.
A larger self-awareness of an individual indicates
a stronger sense of self-protection and a lower willingness of resource donation.
Besides, we also consider that the susceptible individuals can perceive
the disease from immediate neighbors.
With the increase of infected neighbors, the susceptible individuals will reduce the probability of resource
donation, since the donation behavior will lead to fewer resources for self-protection and a higher
probability of been infected.
Then the interplay between resource allocation and epidemic spreading is studied in our paper.

We first study the effects of degree and self-awareness heterogeneity on epidemic dynamics on the artificial networks.
Through extensive Monte Carlo simulations, we find that the heterogeneity of degree distribution enhances the epidemic spreading, which is consistent with the classical results on scale-free networks \cite{pastor2002epidemic,barthelemy2004velocity}. Besides, we find that the self-awareness heterogeneity suppresses the outbreak of an epidemic. Next, we study the influence of the correlation between node degree and self-awareness on the epidemic dynamics on artificial networks. We find that when there is a positive correlation between node degree and self-awareness, the heterogeneity of self-awareness restrains the outbreak of the epidemic. While, when there is a strong negative correlation, the epidemic threshold first increases and then decreases with the decline of the self-awareness heterogeneity.
Furthermore, we find an optimal self-awareness distribution, at which the disease can be suppressed to the most extent. By exploring the relationship between the epidemic threshold and correlation coefficient, we reveal that the more positive correlation between node degree and self-awareness, the better the disease can be suppressed. Besides, a critical value of the correlation coefficient is found. When the coefficient is below the critical value, an optimal heterogeneity of self-awareness exists.
In contrast, the epidemic threshold decreases monotonously with the decline of the self-awareness heterogeneity. At last, we verify the results on four typical real-world networks, and
find that the results on the real-world networks are consistent with those on the artificial network.

\section{Model descriptions} \label{sec:model}
In this section, a resource-epidemic coevolution model that is named as the resource-based epidemiological susceptible-infected-susceptible model (r-SIS)
is proposed.

\subsection{Epidemic spreading model}
For the epidemic spreading model, each node has two possible states: the infected (I) and the
susceptible (S) state~\cite{peng2012reaction}. At any time step $t$, each I-state node $i$ can
transmit the disease to its susceptible neighbors, at the
same time it recoveries to S-state with the rate $\mu_i(t)$, which depends on the resource quantity
$\omega_i(t)$ received from its healthy neighbors. We consider that the resource, such as funds and
medical, can promote the recovery of the I-state individuals
\cite{kulik1989social,nausheen2009social}. Thus, the recovery rate $\mu_i(t)$ is assumed to be
proportional to resource quantity $\omega_i(t)$ in this paper and is defined as:
\begin{equation}\label{recRate}
 \mu_i(t)=1-(1-\mu_0)^{\varepsilon{\omega_i(t)}},
\end{equation}
where $\mu_0$ is the spontaneous recovery rate that is independent of resource. Since the value of $\mu_0$ does not qualitatively affect the results \cite{chen2018controlling}, without loss of generality, it is set at a small value $\mu_0=0.1$ in this paper. Besides, resource wastes widely exist in the real scene of the medical system during the treatment process \cite{mackie2007health}. To mimic the phenomenon of resource wasting, a parameter $\varepsilon$ is introduced to represent the resource utilization rate, which has been proved to not qualitatively affect the dynamical properties \cite{chen2018suppressing,chen2018controlling}.
Thus, without loss of generality, it is set to be $\varepsilon=0.6$ in this paper.
In the spreading process of the epidemic, every S-state node has a probability of being infected by
its I-state neighbors. We consider that the healthy (S-state) nodes are the source of resources,
they can both generate new resources and donate them to the I-state neighbors. If an S-state
node chooses to donate its resources, it will have fewer resources for self-protection, which leads
to a greater risk of been infected. Otherwise, if it refuses to donate resources for self-protection,
the infection rate will reduce by a factor $c$. As there is a heterogeneous distribution of
self-awareness in populations, the infection rate varies from node to node. To reflect the
relationship between the donation behavior and infection probability, we denote the basic
infection rate as $\beta$, and define the actual infection of node
$i$ as:
\begin{equation}\label{lambda}
\beta_i(t)=
\left\{
     \begin{array}{ll}
       \beta, & \hbox{if donating resource;} \\
       c\beta,& \hbox{otherwise. }
     \end{array}
   \right.
\end{equation}

According to the individual-based mean-field
theory \cite{sharkey2011deterministic,pastor2015epidemic},
the dynamical process of the r-SIS model can be expressed as:
\begin{equation}\label{eqRho}
\frac{d\rho_i(t)}{dt}=-\mu_i(t)\rho_i(t)+\beta_i(t)[1-\rho_i(t)]\sum_{j=1}^{N}a_{ij}\rho_j(t),
\end{equation}
where $a_{ij}$ is the element of adjacency matrix $A$. If there is an
edge between nodes $i$ and $j$, $a_{ij}=1$, otherwise, $a_{ij}=0$.
Besides, to calculate the spreading size of the epidemic,
a parameter $\rho_i(t)$ is introduced to represent the probability
that node $i$ is in I-state. Thus we can calculate the fraction of infected
nodes in a network of size $N$ at time $t$ by
averaging overall $N$ nodes:
\begin{equation}
  \rho(t)=\frac{1}{N}\sum_{i=1}^{N}\rho_i(t).
  \label{density}
\end{equation}
At last, the prevalence of the epidemic in the stationary state is defined as
$\rho\equiv\rho(\infty)$.
\subsection{Resource allocation model}\label{sec:res}
As the statements above, the healthy individuals can generate and donate
resources to support the recovery of their I-state neighbors during
the outbreak of an epidemic. For the sake of simplicity, we assume
that each susceptible node will generate one unit resource at each time
step. Besides, the S-state nodes can perceive
the severity of an outbreak from the state of infection in the neighborhood.
The parameter $m_i$ is defined to represent the amount of the I-state
neighbors of node $i$. Generally, the larger the value of $m_i$,
the lower probability of resource donation of node $i$ \cite{funk2010modelling,wang2015dynamics}.
Besides, the parameter $\alpha_i$ is defined to
represent the awareness of self-protection for each node $i$.
A larger value of $\alpha_i$ indicates more sensitive
of node $i$ to the disease, and a lower intention to
donate resource. We consider that $\alpha_i$
obeys the heterogeneous distribution $g(\alpha)\sim{\alpha}^{-\gamma_{\alpha}}$,
where $\gamma_{\alpha}$ is the self-awareness exponent.
Thus, the resource donation probability of a healthy node $i$ is related
to its intrinsic self-awareness and a total of $m_i$ infected neighbors,
which is defined as:
\begin{equation}\label{disProb}
  q_i(t)=q_0(1-\alpha_i)^{m_i(t)},
\end{equation}
where $q_0$ is a basic donation probability. We consider
that the resource of an S-state node will be
distributed equally among neighbors.

Based on the above resource allocation scheme,
the amount of resource that node $i$ donates to node $j$
at a time can be expressed as:
\begin{equation}
  \omega_{i\rightarrow j}(t)=q_i(t)\frac{1}{m_i(t)}.
  \label{ResTrans}
\end{equation}
With the resource allocation scheme defined in Eq.~(\ref{ResTrans}),
we can further define the resource quantity $\omega_{i}(t)$,
which is expressed as:
\begin{equation}
\begin{split}
\omega_i(t)&=\sum_{j\in \mathcal{N}_{i}}[1-\rho_j(t)]\omega_{j\rightarrow i}(t)\\
&=\sum_{j\in \mathcal{N}_{i}}[1-\rho_j(t)]\frac{q_j(t)}{m_j(t)}.
\label{resource}
\end{split}
\end{equation}
where $\mathcal{N}_{i}$ represents the neighbor set of node $i$, and
the expression $[1-\rho_j(t)]$ represents the probability that
a neighbor $j$ is in S-state.
Combining Eq.(~\ref{ResTrans}), we can
get the expression of infection rate of any node $i$ at time $t$ as:
\begin{equation}\label{lami}
  \beta_i(t)=q_i(t)\beta+[1-q_i(t)]c\beta
\end{equation}

\section{Simulation results}
Although various theoretical methods such as the heterogeneous mean-field (HMF), quench mean-field (QMF) and dynamical message passing (DMP) approaches
have been proposed to analyze both the single dynamical process \cite{wang2016unification}
and the multiple coupled dynamical processes \cite{kabir2020impact,kuga2018impact}, the nonlinearity
of the model described in section~\ref{sec:model} and the strong dynamic correlations
make it infeasible to obtain precise theoretical solutions for epidemic size and threshold by utilizing the existing theoretical methods. Therefore,
extensive Monte Carlo simulations are carried out to study the coevolution
of resource allocation and epidemic spreading in this section.
First, we study the impact of heterogeneous distributions of self-awareness
and node degree on the epidemic dynamics, and then investigate the effects of
correlation between node degree and self-awareness on the spreading dynamics on artificial networks
through Monte Carlo simulations. At last, we will verify the results by conducting
simulations on several typical real-world networks.

A specific simulation is carried out as
follows~\cite{schonfisch1999synchronous}:
during each time time interval $[t, t+\Delta{t}]$, each S-state node $i$ changes to
I-state in rate $\beta_i$, which can be defined as~\cite{fennell2016limitations}
\begin{equation}\label{infRate}
  \beta_i(t)={\rm lim}_{\Delta{t}\rightarrow{0}}\frac{P(S_{t+\Delta{t}}^{i}=I {\rm ~infected~by~} {j}|S_{t}^{i}=S,S_{t}^{j}=I)}{\Delta{t}},
\end{equation}
where $S_{t}^{i}$ is denoted as the state of node $i$ at time $t$, and $(S_{t+\Delta{t}}^{i}=I {\rm~infected~by~} {j})$
represents that node $i$ is infected by an I-state neighbor $j$ \cite{cai2019precisely}. At the same time, the I-state node will recover to S-state with rate of $\mu_{i}(t)$, which defines as
\begin{equation}\label{recRate}
  \mu_i(t)={\rm lim}_{\Delta{t}\rightarrow{0}}\frac{P(S_{t+\Delta{t}}^{i}=S |S_{t}^{i}=I)}{\Delta{t}}.
\end{equation}
The infection rate $\beta_j(t)$ and recovery rate $\mu_i(t)$ are
dependent on the resource donation probability $q_j(t)$ and resource quantity $\omega_i(t)$ respectively.
The process of resource allocation takes place with the propagation of disease.
In synchronous updating, the $\Delta{t}$ is finite, and the infection and recovery
probability of node $i$ is $\tilde{\beta_i}=\beta_i\Delta{t}$, and $\tilde{\mu}_i=\mu_i{\Delta{t}}$.
According to Eqs.(\ref{infRate}) and (\ref{recRate}), the transition probabilities can be expressed as:
\begin{equation}\label{infProb}
  \beta_i{\Delta{t}}=P(S_{t+\Delta{t}}^{i}=I {\rm~infected~by~} j|S_{t}^{i}=S,S_{t}^{j}=I),
\end{equation}
\begin{equation}\label{recProb}
  \mu_i\Delta{t}=P(S_{t+\Delta{t}}^{i}=S|S_{t}^{i}=I).
\end{equation}
At the end of each time step, the state of all nodes in the network update synchronously.
To ensure that the dynamical processes enter a stationary,
in which the prevalence fluctuates within a small range,
each simulation will run a sufficiently long time. Besides, in
order to avoid the influence of other factors on the results,
without loss of generality, we set the coefficient
$c$ at a constant value $c=0.05$, such that if any healthy individual $j$
chooses to reserve its resource, the probability that it is
infected in one contact with an infected neighbor
reduces to $\beta_j=0.05\beta$.

\subsection{Effects of heterogeneous self-awareness and degree distributions }\label{sec:heterDist}
In order to investigate the effects of heterogeneous distributions of self-awareness and
node degree on the epidemic dynamics, we first generate networks with
degree distribution $P(k)\sim{k^{-\gamma_D}}$ using the uncorrelated configuration model
(UCM) \cite{molloy1995critical,catanzaro2005generation}, where $\gamma_D$ is the degree exponent.
The maximum and minimum degree of the network are set to be $k_{\rm max}\sim\sqrt{N}$ and $k_{\rm min}=3$ respectively,
which assures no degree correlation of the network when $N$ is sufficient large
\cite{boguna2004cut,clauset2009power}, and the mean degree is set to be
$\langle{k}\rangle=8$. Then we generate a self-awareness sequence $\{\alpha_i\}_{i=1}^{N}$
according to the distribution $g(\alpha)\sim{\alpha}^{-\gamma_{\alpha}}$. The maximum and minimum
values are set to be $\alpha_{\rm max}\sim{N}^{1/2}$, $\alpha_{\rm min}=5$ respectively.
To ensure $\alpha\in[0,1]$, we rescale each value as $\alpha/\alpha_{\rm max}$. At last, each
node is assigned an value of self-awareness randomly.

In addition, we employ the susceptibility measure ~\cite{ferreira2012epidemic}
$\chi$ to numerically determine the epidemic threshold:
\begin{equation}
\chi = N\frac{\langle\rho^2\rangle-\langle\rho\rangle^2}{\langle\rho\rangle},
\end{equation}
where $\langle\cdots\rangle$ is the ensemble averaging. To obtain a reliable
value of $\chi$, we perform at least $2\times10^3$ independent realizations on
a specific network with fixed self-awareness distribution for each basic infection
rate $\beta$. At the threshold $\beta_c$, the value of $\chi$ exhibits a maximum value.
And then, by performing the simulations on 100 different networks, we can obtain the
average value of $\beta_c$.
\begin{figure}
  \centering
  \includegraphics[width=0.7\linewidth]{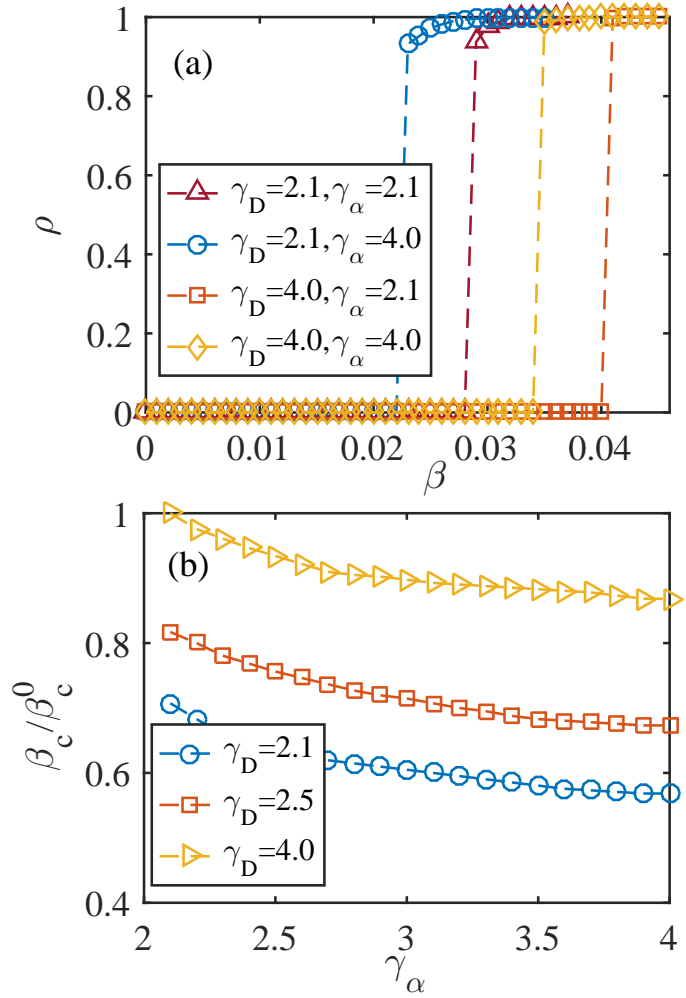}\\
  \caption{The influence of degree and awareness distributions on the spreading dynamics without
  correlations. (a) The prevalence $\rho$ in the stationary state versus the basic infection rate $\beta$
  for $\gamma_D=2.1$, $\gamma_{\alpha}=2.1$ (red up-triangles), $\gamma_D=2.1$, $\gamma_{\alpha}=4.0$
  (blue circles), $\gamma_D=4.0$, $\gamma_{\alpha}=2.1$ (orange squares), and $\gamma_D=4.0$, $\gamma_{\alpha}=4.0$
  (yellow rhombus) respectively. (b) The relative epidemic threshold $\beta_c/\beta^0_c$ as a function of
  awareness exponent $\gamma_{\alpha}$ for degree exponent $\gamma_D=2.1$ (blue circles), $\gamma_D=2.5$ (orange squares),
  and $\gamma_D=4.0$ (yellow right-triangles), where $\beta^0_c\approx0.041$ is the threshold
  at $\gamma_D=4.0,\gamma_{\alpha}=2.1$. Data are obtained by averaging over 500 independent simulations.}
  \label{infsize_uncorr}
\end{figure}

Fig.~\ref{infsize_uncorr} (a) displays the prevalence $\rho$ in the stationary state
as a function of basic infection rate $\beta$ at different degree and awareness exponents.
We can observe that the dynamic process converges to two possible stationary states:
the completely healthy state when $\beta<\beta_c$, and nearly all infected state $\beta>\beta_c$,
which indicates a first-order transition at $\beta_c$. As shown in Fig.~\ref{infsize_uncorr} (a),
the epidemic threshold $\beta_c$ decreases with the heterogeneity of degree distribution, which is consistent with the epidemic outbreak on networks without self-awareness \cite{pastor2001epidemicprl}. The phenomenon is induced by the hub nodes that exist on strong heterogeneous networks. When a given heterogeneity of degree distribution, the value of $\beta_c$ increases with the heterogeneity of self-awareness distribution, which is in contrast to the effects of degree heterogeneity. For instance, when $\gamma_D=2.1$, the $\beta_c$ for $\gamma_{\alpha}=2.1$ is larger than that for
$\gamma_{\alpha}=4.0$.  Fig~\ref{infsize_uncorr} (b) exhibits the value of relative
threshold $\beta_c/\beta_c^0$ as a function of $\gamma_{\alpha}$ for three
degree exponents $\gamma_D=2.1$, $\gamma_D=2.5$ and $\gamma_D=4.0$ respectively.
We can observe that the epidemic threshold $\beta_c$ decreases gradually
with the increases of $\gamma_{\alpha}$, which reveals that the self-awareness
heterogeneity restrains the epidemic spreading.

\begin{figure}
  \centering
  \includegraphics[width=0.9\linewidth]{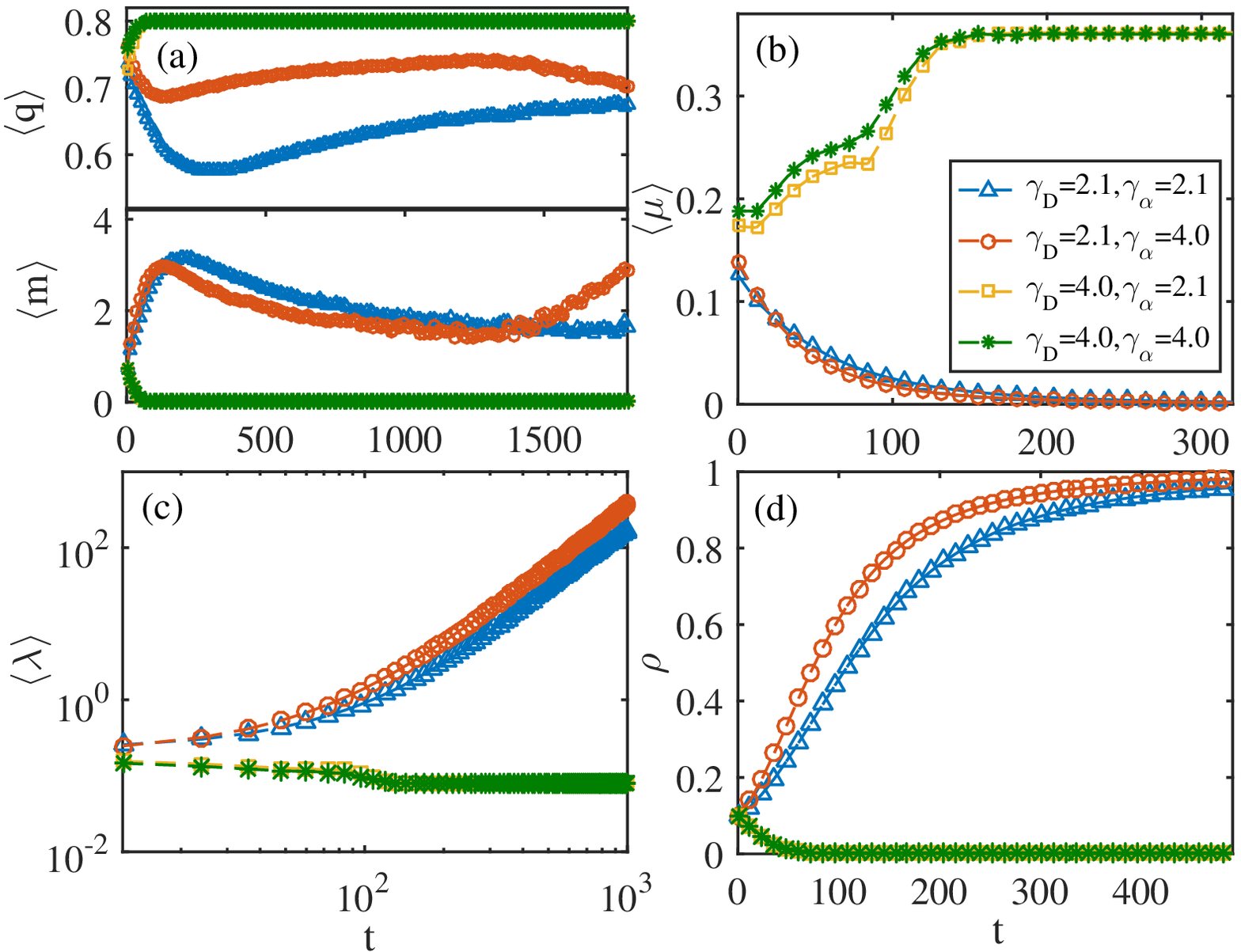}\\
  \caption{Plots of the critical parameters versus time $t$. (a) Top pane: time evolution of
  the average resource donation probability $\langle{q}\rangle$ for
  $\gamma_D=2.1$, $\gamma_{\alpha}=2.1$ (blue up-triangles), $\gamma_D=2.1$, $\gamma_{\alpha}=4.0$
  (orange circles), $\gamma_D=4.0$, $\gamma_{\alpha}=2.1$ (yellow squares), and $\gamma_D=4.0$, $\gamma_{\alpha}=4.0$
  (green snowflakes) respectively; Bottom pane: the corresponding time evolution of the average number of
  infected neighbors $\langle{m}\rangle$. (b) The average recovery rate $\langle{\mu}\rangle$ versus $t$.
  (c) The time evolution of average effective infection rate $\langle{\lambda}\rangle$. (d) The time evolution
  of the fraction of infected nodes $\rho(t)$.  Data are obtained by averaging over 500 independent Monte Carlo simulations.
  }
  \label{time_uncorr}
\end{figure}

Next, we explain qualitatively the results above by exploring the time evolutions of the critical parameters when the basic infection rate is set to be $\beta=0.04$. Fig~\ref{time_uncorr} (a) plots the time evolution of the
average donation probability $\langle{q}\rangle$ (top pane) and average number of infected neighbors $\langle{m}\rangle$ for varies values of $\gamma_D$ and $\gamma_{\alpha}$ (bottom panel). It shows that when $\gamma_D=2.1$, the donation probability $\langle{q}\rangle$ first decreases and then increases with time $t$ for $\gamma_{\alpha}=2.1$ (see blue upper-circles). While, for $\gamma_{\alpha}=4.0$ (see red circles), the value of $\langle{q}\rangle$ first decreases, and then increases slightly in the middle time, at last, it declines with time $t$. This phenomenon can be qualitatively explained as follows: When there is a strong heterogeneity of the self-awareness distribution, for instance, $\gamma_{\alpha}=2.1$, the network has a larger number of nodes with very low self-awareness, and more nodes with high self-awareness. Since the more nodes with large self-awareness means a lower willingness to donation resources, and a lower donation probability
in the early stage, thus the value of $\langle{q}\rangle$ drops more abruptly than that for $\gamma_{\alpha}=4.0$, which induces a larger number of infected neighbors $\langle{m}\rangle$, as less resource is donated from healthy nodes to their neighbors. This phenomenon can be verified by studying the bottom pane of Fig.~\ref{time_uncorr} (a). With less resource, the I-state nodes will recovery with a relatively lower recovery rate [see blue upper-triangles and red circles in Fig.~\ref{time_uncorr}(b)]. With the spread of the disease, more nodes with a very small value of $\alpha$ participate in the behavior of
donating resources. The smaller value of $\alpha$ means a larger willingness to donate resource, which leads to a greater growth of $\langle{q}\rangle$ for $\gamma_{\alpha}=2.1$
that $\gamma_{\alpha}=4.0$ [see the blue upper-triangles and red circles in the top pane of Fig~\ref{time_uncorr} (a)]. Consequently, with the increase of donation probability of the entire network, the value of $\langle{m}\rangle$ decreases gradually, which leading
to a slower decrease of $\langle{\mu}\rangle$ for $\gamma_{\alpha}=2.1$
that $\gamma_{\alpha}=4.0$ [see the blue upper-triangles and red circles
in Fig~\ref{time_uncorr} (b)]. The relative larger recovery rate leads
to a lower effective infection rate, which is defined as $\langle{\lambda}\rangle=\langle\beta\rangle/\langle{\mu}\rangle$, as shown in Fig.~\ref{time_uncorr} (c). Thus the prevalence $\rho(t)$ increases slower for $\gamma_{\alpha}=2.1$ than $\gamma_{\alpha}=4.0$, as shown in Fig~\ref{time_uncorr} (d).
From the statement above, we can explain the reason why the heterogeneity of self-awareness distribution can suppress the outbreak of an epidemic.

When the degree heterogeneity of the network decreases, for instance, $\gamma_D=4.0$, the donation probability for both $\langle{q}\rangle$ for $\gamma_{\alpha}=2.1$ that $\gamma_{\alpha}=4.0$ increases with time $t$ [see the yellow squares and green snowflakes in the top pane of
Fig~\ref{time_uncorr} (a)], which leads to an increase of the recovery rate of the whole network, as shown in Fig.~\ref{time_uncorr} (b). Consequently, the effective infection rate of the network $\langle{\lambda}\rangle$ decreases gradually to a very small value with time $t$. Thus we  see that the prevalence decreases gradually to zero [see Fig~\ref{time_uncorr} (d)].
From the statement above, we can also explain the phenomenon that the
threshold $\beta_c$ decreases with an increase of $\lambda_D$.

\subsection{Effects of degree-awareness correlations }\label{sec:corr}
In this section, we focus on how the correlation between
the node degree and self-awareness affects the spreading dynamics.
A network with a given correlation coefficient is built as follows:
\begin{itemize}
  \item A network with degree distribution $P(k)\sim{k}^{-\gamma_D}$ is built
by the steps in \emph{Section~\ref{sec:heterDist}};
  \item A self-awareness sequence is generated from the distribution
$P(\alpha)\sim\alpha^{-\gamma_{\alpha}}$;
  \item Sorting the nodes of the network by the degree in ascending order,
and then sorting the self-awareness sequence in ascending or descending order, respectively.
  \item Rearranging the order of each self-awareness value with a given probability
$\pi$, and then assigning each self-awareness value to the corresponding node.
\end{itemize}
According to the steps above, we can get a network with a given
awareness-degree correlation. The correlation coefficient is:
\begin{equation}\label{sigma}
  \sigma=1-\pi.
\end{equation}
If the self-awareness sequence is sorted in ascending order, we can get a
positive awareness-degree correlation with coefficient $\sigma$;
otherwise, we can get a
negative awareness-degree correlation with coefficient $\sigma$.

\begin{figure}
  \centering
  \includegraphics[width=1.0\linewidth]{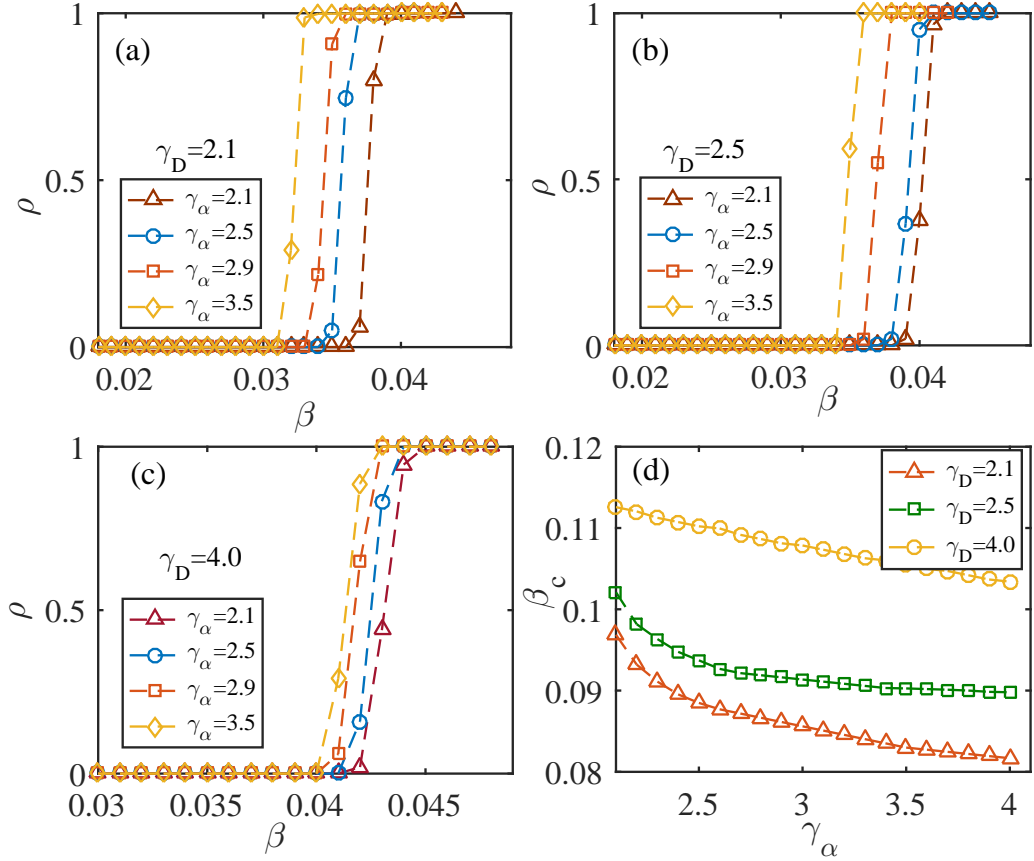}\\
  \caption{Impacts of self-awareness heterogeneity on spreading dynamics with positive correlation
  between node degree and self-awareness. (a) The prevalence $\rho$ in the stationary
  state as a function of basic infection rate $\beta$ for $\gamma_{\alpha}=2.1$ (red upper-triangles), $\gamma_{\alpha}=2.5$ (blue circles), $\gamma_{\alpha}=2.9$ (orange squares),
  and $\gamma_{\alpha}=3.5$ (yellow rhombuses) respectively when degree exponent is fixed at
  $\gamma_D=2.1$. (b) The value of $\rho$ versus $\beta$ for the corresponding $\gamma_{\alpha}$ when $\gamma_D=2.5$. (c) The value of $\rho$ as a function of $\beta$
  for the corresponding $\gamma_{\alpha}$ when $\gamma_D=4.0$. (d) The
  epidemic threshold $\beta_c$ as a function of $\gamma_{\alpha}$ for $\gamma_D=2.1$ (orange upper-triangles), $\gamma_D=2.5$ (green squares), and $\gamma_D=4.0$ (yellow circles) respectively.
  The correlation coefficient is $\sigma=0.8$. Data are obtained by averaging over 500 independent Monte Carlo simulations.
   }\label{infcorr_0.8}
\end{figure}

First of all, we study the case when there is a positive degree-awareness correlation, i.e., $\sigma=0.8$. Figs.~\ref{infcorr_0.8} (a), (b) and (c) exhibit the prevalence $\rho$ in the
stationary as a function of $\beta$ when degree exponent is $\gamma_D=2.1$, $\gamma_D=2.5$
and $\gamma_D=4.0$, respectively. It shows that for a network with a fixed structure, the epidemic threshold $\beta_c$ increases with self-awareness heterogeneity.
To verify the results obtained in Figs.~\ref{infcorr_0.8} (a) to (c), we
explore the relationship between $\beta_c$ and the awareness exponent $\gamma_{\alpha}$ in Fig.~\ref{infcorr_0.8} (d). Obviously, for each fixed $\gamma_D$, the value of $\beta_c$ decreases gradually with the $\gamma_{\alpha}$, which suggests that
when node degree and self-awareness correlated positively, the heterogeneity of self-awareness inhibits the epidemic spreading.

\begin{figure}
  \centering
  \includegraphics[width=1.0\linewidth]{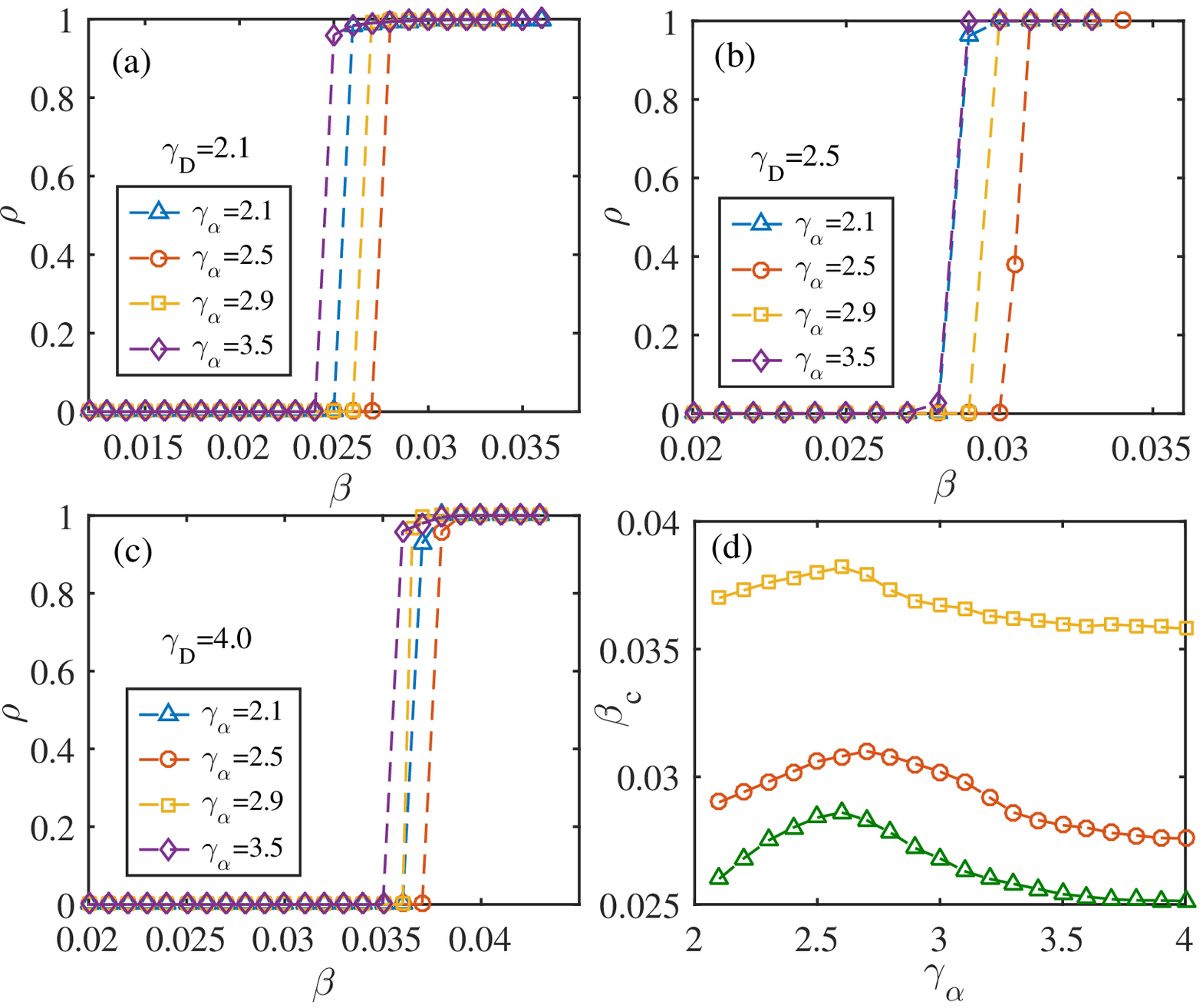}\\
  \caption{The influence of self-awareness heterogeneity on spreading dynamics with negative correlation
  between node degree and self-awareness. (a) The prevalence $\rho$ in the stationary
  state as a function of basic infection rate $\beta$ for $\gamma_{\alpha}=2.1$ (blue upper-triangles), $\gamma_{\alpha}=2.5$ (orange circles), $\gamma_{\alpha}=2.9$ (yellow squares),
  and $\gamma_{\alpha}=3.5$ (purple rhombuses) respectively when degree exponent is fixed at
  $\gamma_D=2.1$. (b) The value of $\rho$ versus $\beta$ for the corresponding $\gamma_{\alpha}$ when $\gamma_D=2.5$. (c) The value of $\rho$ as a function of $\beta$
  for the corresponding $\gamma_{\alpha}$ when $\gamma_D=4.0$. (d) The
  epidemic threshold $\beta_c$ as a function of $\gamma_{\alpha}$ for $\gamma_D=2.1$ (green upper-triangles), $\gamma_D=2.5$ (orange cirlces), and $\gamma_D=4.0$ (yellow squares) respectively.
  The correlation coefficient is $\sigma=-0.8$. Data are obtained by averaging over 500 independent Monte Carlo simulations.}\label{infcorr_-0.8}
\end{figure}
Next, we explore the case when there is a negative degree-awareness correlation.
Figs.~\ref{infcorr_-0.8} (a) to (c) display the value of $\rho$ versus $\beta$ for
$\gamma_D=2.1$, $\gamma_D=2.5$ and $\gamma_D=4.0$ when $\sigma=-0.8$ respectively.
It shows that for a fixed network, for instance $\gamma_D=2.1$, the epidemic threshold
$\beta_c$ first increases when $\gamma_{\alpha}$ increase from $\gamma_{\alpha}=2.1$ to
$\gamma_{\alpha}=2.5$, and then it decreases when $\gamma_{\alpha}$ increases from
$\gamma_{\alpha}=2.5$ to $\gamma_{\alpha}=4.0$. We next study systematically
the effects of negative correlation between node degree and self-awareness on
the spreading dynamics by exploring the relationship between threshold $\beta_c$
and awareness exponent $\gamma_{\alpha}$ in Fig.~\ref{infcorr_-0.8} (d) for
$\gamma_D=2.1$ (green upper-triangles), $\gamma_D=2.5$ (red circles), and
$\gamma_D=4.0$ (yellow squares) respectively.
It shows that the threshold $\beta_c$ first
increases and then decreases with $\gamma_{\alpha}$, and an optimal
value $\gamma_{\alpha}^{\rm opt}$ ($\gamma_{\alpha}^{\rm opt}$ is around $2.6$) exists,
at which the value of $\beta_c$ reaches maximum. This result can be qualitatively explained as follow: When the self-awareness heterogeneity is very strong, e.g., $\gamma_{\alpha}=2.1$, there is a large number of nodes with very small values of self-awareness $\alpha$ (strong willingness of resource donation),
and many nodes with very large values $\alpha$ (weak willingness of resource donation).
When the coefficient $\sigma=-0.8$, there is a strong negative correlation
between node degree and self-awareness. Under this circumstance,
the large degree nodes will have a strong willingness to donate resource,
while the small degree nodes (covering most nodes of the network)
have a weak willingness to donate resource,
which leads to a high infection rate of these hub nodes.
Besides, the epidemic spreading dynamics exhibits hierarchical features
\cite{barthelemy2004velocity}. That is to say,
the hubs with large degrees are more likely to be infected firstly,
and then the disease propagates from hubs to the intermediate nodes,
and finally to nodes with small degrees. Therefore, the large numbers of
small degrees will be infected rapidly by the hub nodes in this situation, and the epidemic will outbreak easily.

When the heterogeneity of self-awareness distribution is weak, i.e.,
$\gamma_{\alpha}=4.0$, there is a small fraction of nodes with very
large or small value of $\alpha$, many nodes have an intermediate $\alpha$
around the mean value $\langle{\alpha}\rangle$. The awareness level
of the small degree nodes reduces compared to the case of $\gamma_{\alpha}=2.1$,
which leads to a raise of both the donation probability $\langle{q}\rangle$ and
effective infection rate $\langle{\lambda}\rangle$ of these nodes.
As the strong negative correlation between node degree and self-awareness, the hub nodes still have a very small value of $\alpha$ (large value of donation probability $\langle{q}\rangle$). During the outbreak of an epidemic, these hubs nodes are more likely to be infected in the early
stage, and then transmit the disease to those small degree nodes rapidly as they have a highly effective infection rate $\langle{\lambda}\rangle$.
Thus the epidemic will break out more easily than the case of $\gamma_{\alpha}=2.1$
in this situation.

According to the above statement,
we have qualitatively explained the optimal phenomenon
by explaining why diseases are more likely to break out when
there is a strong heterogeneity ($\gamma_{\alpha}=2.1$) and
weak heterogeneity ($\gamma_{\alpha}=4.0$).

\begin{figure}
  \centering
  \includegraphics[width=0.7\linewidth]{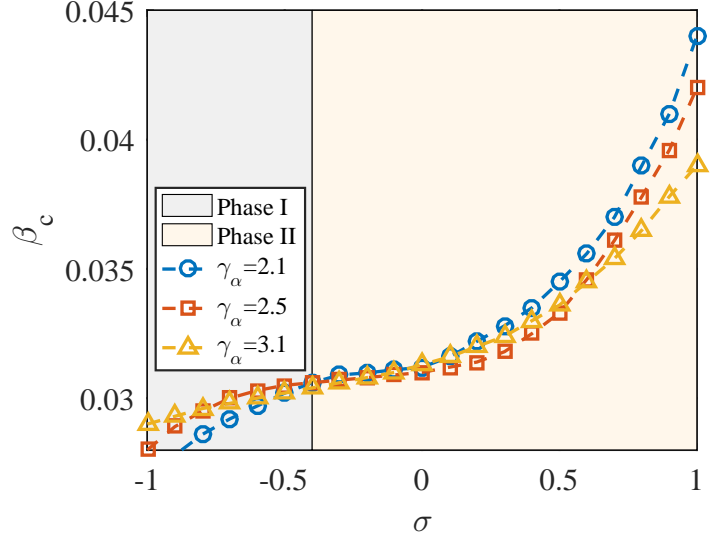}\\
  \caption{The epidemic threshold $\beta_c$ as a function of correlation coefficient
  $\sigma$ for $\gamma_{\alpha}=2.1$ (blue circles), $\gamma_{\alpha}=2.5$ (red squares), and
  $\gamma_{\alpha}=3.1$ (yellow upper-triangles). The degree exponent is fixed at $\gamma_D=2.5$.
  Phase $I$ and phase $II$ are separated by critical value $\sigma_c\approx-0.4$. Data are obtained by averaging over 500 independent Monte Carlo simulations.}
  \label{sigma_threshold}
\end{figure}

Next, we study the effects of correlation between node degree and
self-awareness by exploring the relationship between epidemic threshold $\beta_c$ systematically
and correlation coefficient. Fig.~\ref{sigma_threshold} displays the
$\beta_c$ as a function of $\sigma$ for $\gamma_{\alpha}=2.1$, $\gamma_{\alpha}=2.5$,
$\gamma_{\alpha}=3.1$ respectively. We find that for each fixed heterogeneity of self-awareness,
the epidemic threshold $\beta_c$ increases monotonously with correlation
coefficient $\sigma$. For instance, for $\gamma_{\alpha}=2.1$, the threshold increases
from $\beta_c\approx0.027$ to $\beta_c\approx0.044$, which suggests that
the more positive of the correlation between node degree and self-awareness, the better the disease can be suppressed.
This phenomenon can be qualitatively explained as follows:
When there is a larger positive correlation, the hub nodes will have a
larger value of $\alpha$, which indicates a smaller probability of resource
donation $\langle{q}\rangle$, and consequently a lower effective infection rate
$\langle{\lambda}\rangle$ of the hubs. This phenomenon reduces the infection probability of the hub nodes in the early stage. Meanwhile, the small degree nodes have a larger probability of resource donation, which increases the recovery rate of
the I-state neighbors, including the hub nodes. Thus the outbreak of the epidemic is
effectively delayed.
In addition, it also shows that the parameter pane $(\sigma-\beta_c)$
is separated into two phases: phase I and phase II, by a critical value of $\sigma\approx0.4$.
In phase $I$, the threshold $\beta_c$ first increases and then decreases with
the increase of $\gamma_{\alpha}$, namely the optimal value $\gamma_{\alpha}^{\rm opt}$ exists in this region, as shown the curves in Fig.~\ref{infcorr_-0.8} (d) for $\sigma=-0.8$.
In phase $II$, the threshold $\beta_c$
decreases monotonously with $\gamma_{\alpha}$, as shown
the curves in Fig.~\ref{infsize_uncorr} (b) for $\sigma=0$, and the
curves Fig.~\ref{infcorr_0.8} (d) for $\sigma=0.8$ respectively.

\subsection{Verification on real-world networks}
In this section, we verify the results obtained on artificial networks
by conducting the simulations on the real-world networks.
The following four typical real-world networks are chosen in our paper:
(i). The \emph{OpenFlights} network \cite{konect:opsahl2010b}.
This network describes the flights between airports in the world.
The nodes represent a portion of the world's airports, and edges represent flights from one airport
to another. There are $N=2939$ nodes and $V_e=30501$ edges in the network with maximum degree
$k_{\rm max}=473$ \cite{openflights}.
(ii). The \emph{Euroroad} network \cite{vsubelj2011robust} is a international road network.
Most road in the network is located in Europe. The nodes of the network represent cities, and the edge between
two nodes denotes that an E-road connects them. There are $N=1174$ nodes and $V_e=1417$
edges with $k_{\rm max}=10$.
(iii). The \emph{face-to-face contact} network \cite{isella2011s}. Nodes in this network represent
the individuals who attended the exhibition \emph{INFECTIOUS: STAY AWAY} in 2009 at the Science Gallery in Dublin, edges describe the face-to-face contacts that were active for at least 20 seconds among individuals during the exhibition \cite{sociopatterns}. There are a total numbers of $N=410$ nodes and $V_e=17298$ edges (contacts) in the network. The maximum degree is $k_{\rm max}=294$.
(iv) The \emph{Facebook} network \cite{leskovec2012learning}. This dataset consists of 'circles' (or 'friends lists') from Facebook. Facebook data was collected from survey participants using this Facebook app. The dataset includes node features (profiles), circles, and ego networks.
There are $N=4039$ nodes and $V_e=88234$ edges in the network.

\begin{figure}
  \centering
  \includegraphics[width=1.0\linewidth]{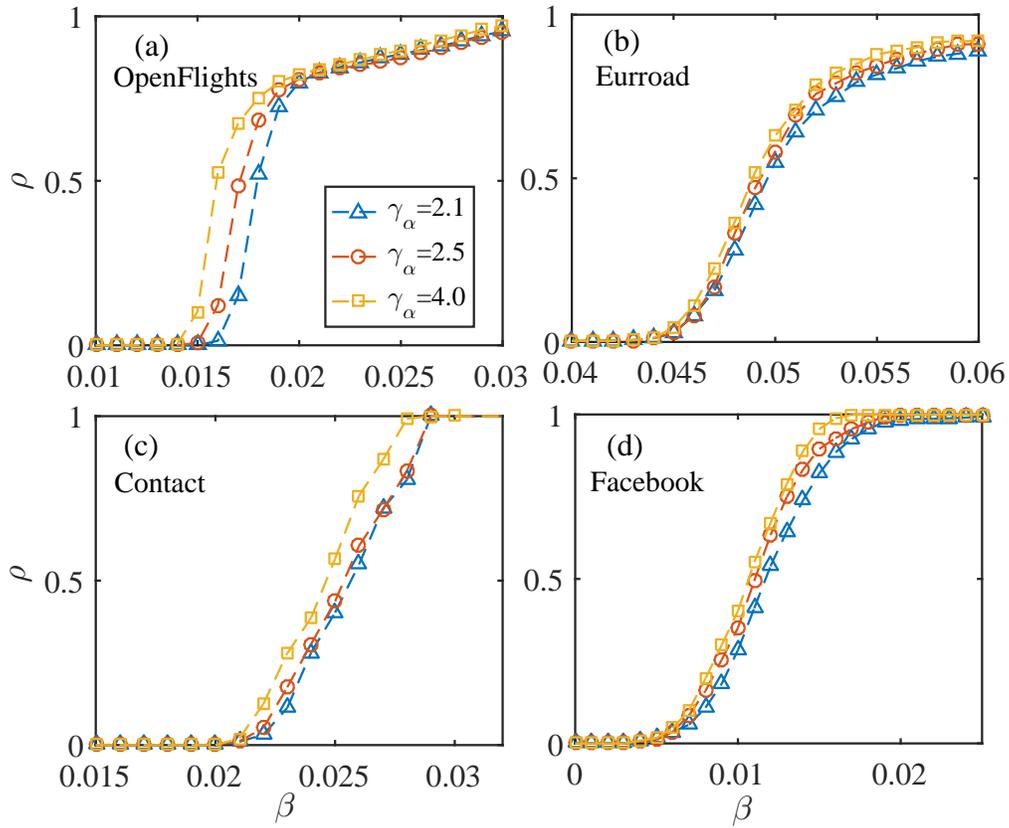}\\
  \caption{The effects of awareness distribution on the epidemic spreading on real
  world networks. (a) The prevalence $\rho$ as a function of basic infection rate $\beta$
  on the \emph{OpenFlights} network for awareness exponent $\gamma_{\alpha}=2.1$ (blue upper-triangle), $\gamma_{\alpha}=2.5$ (red circles),
  and $\gamma_{\alpha}=4.0$ (yellow squares) respectively. (b) The prevalence $\rho$ versus $\beta$ for the corresponding
  values of awareness exponent on the \emph{Euroroad} network. (c) The
  value of $\rho$ as a function of $\beta$ for the four typical values of $\gamma_{\alpha}$
  on the face-to-face contact network. (d) The results on \emph{Facebook} network.
  The correlation coefficient between self-awareness and node degree is $\sigma=0$.}\label{real_omega}
\end{figure}
Figs.~\ref{real_omega} (a) to (d) display the prevalence $\rho$ in stationary state
as a function of $\beta$ when there is no degree-awareness correlation on \emph{OpenFlights} network,
\emph{Euroroad} network, \emph{face-to-face contact} network and the \emph{Facebook} network respectively for different heterogeneities of self-awareness. We find that on the \emph{OpenFlights} network,
the threshold $\beta_c$ decreases with the increase of $\gamma_{\alpha}$, which is consistent with the result on artificial networks [see Fig.~\ref{infsize_uncorr} (a)], and the prevalence $\rho$ increases with $\gamma_{\alpha}$ for a fixed basic infection rate $\beta$. However, there is a difference when disease propagates on the other three real-world networks, i.e., the awareness heterogeneity does not alter the threshold of $\beta_c$. Besides, it shows that the prevalence increases with
the decreases of awareness heterogeneity, which is consistent
with the result on \emph{OpenFlights} network. The results above reveal that
the awareness heterogeneity can suppress the outbreak of epidemic on real-world networks
when there is no correlation between node degree and self-awareness.

\begin{figure}
  \centering
  \includegraphics[width=1.0\linewidth]{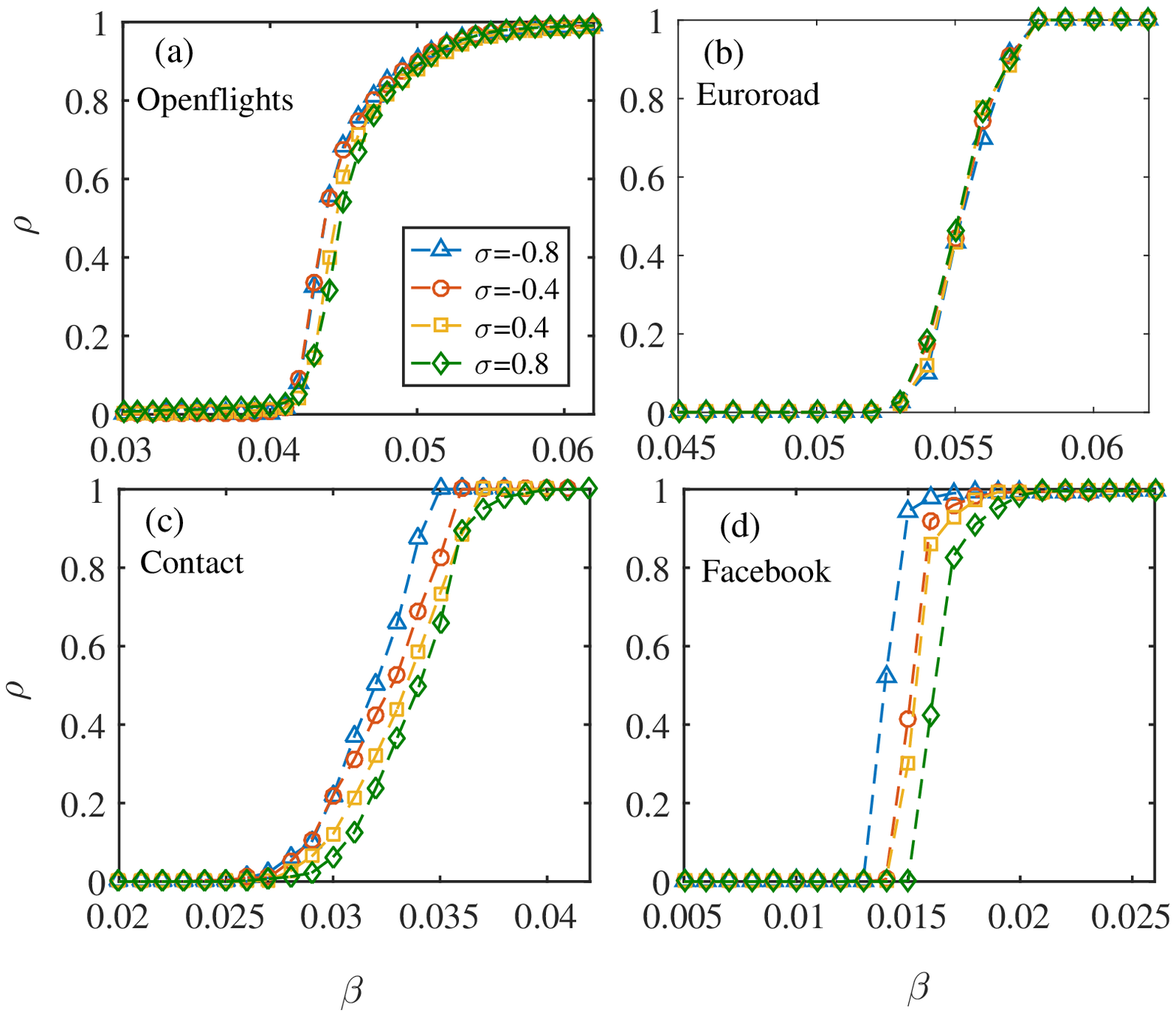}\\
  \caption{The influence of correlation between self-awareness and node degree
   on the epidemic spreading on real world networks. (a) The prevalence $\rho$ as a function of basic infection rate $\beta$
   on the \emph{OpenFlights} network for $\sigma=-0.8$ (blue upper-triangle), $\sigma=-0.4$ (red circles), $\sigma=0.4$ (yellow squares), and $\sigma=0.8$ (green rhombuses) respectively.
   (b) The prevalence $\rho$ versus $\beta$ for the corresponding
  values of $\sigma$ on the \emph{Euroroad} network. (c) The
  value of $\rho$ as a function of $\beta$ for the four typical values of $\sigma$
  on the face-to-face \emph{contact} network. (d) The results on \emph{Facebook} network.
  The awareness exponent is set to be $\gamma_{\alpha}=2.5$ .}
  \label{real_sigma}
\end{figure}
\vspace{10pt}
At last, we study the effects of degree-awareness correlation on the epidemic spreading on the four real-world networks. The awareness exponent is fixed at $\gamma_{\alpha}=2.5$. It shows that when disease propagates on the \emph{OpenFlights} and face-to-face \emph{contact} networks, the threshold $\beta_c$ does not be altered, but the prevalence $\rho$ decreases with correlation coefficient $\sigma$. This result indicates that a more positive correlation between node degree and self-awareness can better suppress the disease spreading, which is consistent with the result of artificial networks. When disease propagates
on the \emph{Euroroad} network, the correlation does not alter both
threshold $\beta_c$ and prevalence $\rho$.
At last, when disease propagates
on the \emph{Facebook} network, the threshold $\beta_c$ increases with $\sigma$, which is consistent with the result on artificial networks [see Fig.~\ref{sigma_threshold}],

Based on the above research, we can see that the results on the real-world networks are consistent with those on the artificial network.
However, the complex structural features of the real networks,
such as clustering, community structure, and small-world characteristics,
have an important impact on the results, and needs to be further
studied in our future research.
\section{Discussion}\label{sec:dis}
In summary, we have studied systematically the impact of heterogeneous awareness
of self-protection on the dynamics of epidemic spreading.
A coevolution dynamical model of resources and epidemic on complex networks has been proposed. The two processes of resource allocation and disease spreading are coevolving in such a way that the generation and allocation of resource depend on the S-state nodes, and the recovery rate of I-state nodes rely on the resources from their S-state neighbors. Both the effective infection rate and the recovery rate of the nodes will be alerted due to the resource factor. First of all, we have
studied the effects of the heterogeneity of both degree and self-awareness distributions
on the epidemic dynamics on artificial networks. Through extensive Monte Carlo simulations, we have found that the degree heterogeneity enhances the epidemic spreading, which is consistent with the classical results on scale-free networks. Besides, the threshold $\beta_c$ increases with the growth of self-awareness heterogeneity for a fixed network structure, which indicates that the self-awareness heterogeneity suppresses the outbreak of an epidemic.

Then we have studied the impact of correlation between node degree and self-awareness on the epidemic dynamics on artificial networks. We have found that when there is a positive correlation between node degree and self-awareness, the threshold $\beta_c$ increases monotonously with the heterogeneity of self-awareness. When there is a strong negative correlation, e.g., $\sigma=-0.8$, the epidemic threshold
$\beta_c$ first increases and then decreases with $\gamma_{\alpha}$. An optimal value $\gamma_{\alpha}^{\rm opt}$ have been found, at which the disease can be suppressed to the most extent. Besides, we have studied systematically the effects of degree-awareness correlation on the epidemic dynamics by exploring the relationship between $\beta_c$ and correlation coefficient $\sigma$. We have found that the epidemic threshold $\beta_c$ increases monotonously with $\sigma$ in $[-1.0,1.0]$, which reveals that the stronger the positive correlation, the more likely the disease can be suppressed. Besides, we have also found a critical value $\sigma_c\approx0.4$, when $\sigma<\sigma_c$, an optimal value $\gamma_{\alpha}^{\rm opt}$ exists, the threshold $\beta_c$ increases with
$\gamma_{\alpha}$ before $\gamma_{\alpha}^{\rm opt}$ and decreases after $\gamma_{\alpha}^{\rm opt}$. When $\sigma>\sigma_c$, $\beta_c$ decreases monotonously with the increase of $\gamma_{\alpha}$.

At last, to verify our results, we have conducted Monte Carlo simulations on four typical real-world networks.
It reveals that the results on the real-world networks are consistent with those on the artificial
network. However, the sophisticated structural features of the real networks, such as clustering, community structure, and small-world characteristics, have an essential impact on the results. For instance, the self-awareness heterogeneity does not alter the threshold $\beta_c$ on the \emph{Euroroad} network, face-to-face contact network, and \emph{Facebook} network. Besides, the correlation does not change both the threshold $\beta_c$ and prevalence $\rho$, when disease propagates
on the \emph{Euroroad} network.

Our findings make a substantial contribution to the understanding of
how the heterogeneous distribution of individual awareness for self-protection influences the dynamics of epidemic spreading. The results in this paper are of practical significance for controlling the outbreak of
infectious diseases, especially in the context of the outbreak of \emph{COVID-19}. It will also guide us to make the most reasonable choice between resource contribution and self-protection when perceiving the threat of disease, and also have a direct application in the
development of strategies to suppress the outbreaks of epidemics.

The present work mainly focus on the spreading dynamics of infectious diseases that can be described by the \emph{susceptible-infected-susceptible} model, such as seasonal influenza.
However, the findings obtained in this paper could still shed light on the control of
the diseases with similar characteristics, such as the SIRS and SIR-like epidemics
that have been widely studied in recent years \cite{arefin2019interplay,kabir2019behavioral,alam2019game,tanimoto2015fundamentals}.
There are still some limitations of our work. For example, as the SIS model
can not describe the spread of an irreversible epidemic,
there would be difference in the
dynamical characteristics such as the transition type and phase diagram between
the SIS model and the irreversible epidemic models. Thus, a coupled dynamic model of resource allocation and epidemic spreading
based on the SIR,SIRV and SEIR models will be studied in our future works. In addition,
the theoretical analysis will also be researched.

\section*{Acknowledgements}
This work was supported by the Fundamental Research Funds for the Central Universities (No. JBK190972), the Financial Intelligence and Financial Engineering Key Laboratory of Sichuan Province,
School of Economic Information Engineering, and the National Natural Science Foundation of China (No. 61673086,61903266), the China Postdoctoral Science Special Foundation (No. 2019T120829), and Sichuan Science and Technology Program (No.20YYJC4001).




\section*{References}

\bibliography{mybibfile}

\end{document}